\documentclass[twocolumn,prb,aps,showpacs]{revtex4}
\usepackage{epsf,latexsym,graphicx}

\newcommand{\eqn}[1]    {(\ref{#1})}

\newcommand{\subbox}[1] {{\mbox{\scriptsize #1}}}

\def\bi         {\begin{itemize}}
\def\ei         {\end{itemize}}
\def\benu   {\begin{enumerate}}
\def\eenu   {\end{enumerate}}
\def\bmat       {\left[ \begin{array}}
\def\emat       {\end{array} \right]}
\def\beq    {\begin{equation}}
\def\eeq    {\end{equation}}
\def\beqn       {\begin{eqnarray*}}
\def\eeqn       {\end{eqnarray*}}
\def\beqa       {\begin{eqnarray}}
\def\eeqa       {\end{eqnarray}}
\def\bquote {\begin{quote}}
\def\equote {\end{quote}}
\def\f          {\frac}
\def\bwide  {\begin{widetext}}
\def\ewide  {\end{widetext}}

\def\d          {\delta}
\def\e          {\epsilon}

\def\g          {\gamma}

\def\m          {\mu}

\def\w          {\omega}

\def\D          {\Delta}

\def\W          {\Omega}

\def\bF         {{\bf F}}

\def\cA     {{\mathcal{A}}}

\def\im     {{\mbox{Im}}}
\def\re     {{\mbox{Re}}}

\def\sgn    {{\mbox{sgn}~}}

\begin{document}

\title{The superconducting critical temperature and
singlet and triplet pair functions of
superconductor/normal-metal/ferromagnet trilayers}
\author{Nayoung Lee}\affiliation{Department of Physics and Institute for
Basic Science Research, SungKyunKwan University, Suwon 440-746,
Korea.}
\author{Han-Yong Choi\footnote{To whom the correspondences should be addressed:
hychoi@skku.edu.} } \affiliation{Department of Physics and Institute
for Basic Science Research, SungKyunKwan University, Suwon 440-746,
Korea.}
\author{Hyeonjin Doh}\affiliation{Department of Physics, University of Toronto,
Toronto, Ontario M5S 1A7, Canada.}
\author{K. Char}
\affiliation{School of Physics and Astronomy and Center for
Strongly Correlated Materials Research, Seoul National
University, Seoul 151-742, Korea.}
\author{Hyun-Woo Lee}
\affiliation{Department of Physics, Pohang University of Science
and Technology, Pohang 790-784, Korea.}

\begin{abstract}
We calculate the superconducting critical temperature $T_c$, the
singlet pair function $\Psi^+(x)$, and triplet pair function
$\Psi^-(x)$ of superconductor/normal metal/ferromagnet (S/N/F)
trilayers using the linearized Usadel equation near $T_c$. The
Green's function method developed by Fominov $et~al.$ for the S/F
bilayers is extended to the S/N/F trilayer systems. The S of the
trilayers is taken to be an $s$-wave singlet pairing
superconductor, and the S/N and N/F interfaces are modeled in
terms of the interface resistances parameterized, respectively,
by $\gamma_b^{SN}$ and $\gamma_b^{NF}$. We present the $T_c$,
$\Psi^+(x)$, and $\Psi^-(x)$ for typical $\gamma_b^{SN}$,
$\gamma_b^{NF}$, and the exchange energy $ E_{ex}$: (a) For a
small (large) $\gamma_b^{NF}$, $T_c$ of S/N/F trilayers, as $d_N$
is increased, increases (decreases) on the length scale of N
coherence length $\xi_N$ with a discontinuity at $d_N=0$ due to a
boundary condition mismatch. (b) $T_c(d_F)$ shows a non-monotonic
behavior like S/F bilayers with a weakened shallow dip. (c) The
odd frequency triplet component $\Psi^-(x)$, induced by $E_{ex}$
and proximity effects, has a maximum near the N/F interface and
decreases on the length scale $\xi_{ex}$ in F. It also penetrates
into N and S regions on the length scale $\xi_N$ and $\xi_S$,
respectively. Based on these results we make comments on the
experimental observation of the odd triplet components and the
recent $T_c$ measurements in Nb/Au/CoFe trilayer systems.
\end{abstract}


\maketitle

\section{Introduction and motivation} \label{sec:introduction}

When different states of matter are in contact, one affects the
other over the coherence length $\xi$, which is referred to as
the proximity
effects.\cite{deGennes64rmp,Deutscher69book,Buzdin05rmp,Bergeret05rmp}
For instance, when a superconductor (S) and a normal metal (N)
are in contact as in an S/N bilayer, the singlet pair function
$\Psi^+ (x) \sim \langle \phi_\uparrow (x) \phi_\downarrow
(x)\rangle$, where $\phi_\sigma (x)$ is the field operator of
spin $\sigma$ electron at the position $x$ and $\langle
~~\rangle$ means the thermodynamic average, penetrates into the N
region and decays exponentially on the length scale of N
coherence length $\xi_N$. For diffusive metals, $\xi_N =
\sqrt{D_N/2\pi T}$ where $D_N$ is the diffusion constant of the
normal metal and $T$ is the
temperature.\cite{deGennes64rmp,Deutscher69book} Note that,
however, the gap energy given by $\Delta(x) = V\Psi^+(x)$
vanishes in N because the pairing interaction $V=0$ in the N
region. The superconducting temperature $T_c$ of an S/N bilayer
as a function of N thickness $d_N$ also decreases exponentially
on the length scale of $\xi_N$. Here, $T_c$ is defined with the
current parallel to the interfaces. Therefore, $T_c$ of an S/N
bilayer is determined by the highest temperature at which the
singlet pair function $\Psi^+(x)$ does not vanish at least at one
point within the bilayer.

The proximity effects in S/F bilayers have also been studied
intensively.\cite{Buzdin05rmp,Bergeret05rmp} Like the S/N case,
$\Psi^+(x)$ of an S/F bilayer decreases exponentially in the F
region, and $T_c$ decreases as the F thickness $d_F$ is increased.
Many of the interesting S/F proximity effects occur at the nanoscale
range of layer thickness. The observation of these effects was made
possible by the progress of fabrication technique of high quality
hybrid S/F layers. The S/F proximity effects have the following
differences compared with S/N: (a) The decay length scale in F,
$\xi_{ex}$, given by $\xi_{ex}=\sqrt{D_F/E_{ex}}$, is much shorter
than $\xi_N$ because the exchange energy $E_{ex}$ is much larger
than $T$. (b) The pair function oscillates as well as decreases in
the F region because the net momentum of a Cooper pair is non-zero
in F.\cite{Demler97prb} By the same reason $T_c(d_F)$ does not
monotonically decrease like S/N as $d_F$ is increased but shows a
shallow dip. The oscillation has been confirmed by the observation
of the ``$\pi$-state'' in S/F/S Josephson
junction,\cite{Kontos01prl,Ryazanov01prl,Kontos02prl,Guichard03prl}
where two superconductors separated by a ferromagnet of appropriate
thickness can have a phase difference of $\pi$. (c) Triplet pairing
component (TPC) is induced in addition to the dominant singlet
pairing component (SPC). The triplet pairing is realized in $^3$He
and Sr$_2$RuO$_4$, for instance, in the form of $p$-wave
pairing.\cite{Maeno03rmp} The induced TPC in S/F systems in the
diffusive limit, however, is even ($s$-wave) in momentum but odd in
frequency, referred to as odd frequency triplet pairing component
(OFTPC). The OFTP was first suggested by Berezinskii in the context
of $^3$He superfluidity,\cite{Berezinskii75jetpl} but it turned out
that it is the $p$-wave triplet pairing that is realized in $^3$He.
The OFTPC is induced from $s$-wave SPC because time reversal
symmetry is broken by the exchange field in S/F bilayers. Just like
non-$s$-wave pairing components are induced from $s$-wave one when
the translational symmetry is broken, OFTPC is induced in S/F
bilayers from the singlet component because the time reversal
symmetry is broken by the exchange field in F region. This can be
seen directly from Fig.\ \ref{fig:sf-PA} where the induced TPC
vanishes as $E_{ex} \rightarrow 0$.

While it is natural to expect that the characteristic length scale
in F region is very short given by $\xi_{ex}$, it was realized
that there exist long range proximity effects in S/F systems set
by the long length scale of $\xi_F=\sqrt{D_F/2\pi T }$. For
instance, the long range triplet pairing was suggested to
understand the unexpected conductance increase of S/F structures
below $T_c$.\cite{Petrashov99prl,Giroud98prb,Bergeret01prl} The
condition for the occurrence of the long length TPC is that the
magnetization is $not$ unidirectional. When the magnetization in F
is unidirectional, the induced triplet pairing component has the
short coherence length $\xi_{ex}$ like the singlet one and can be
represented as $S_z = 0$ component along the magnetization
direction in F.\cite{Eschrig05prb} More explicitly, it is given by
$\langle\phi_\uparrow(x,\tau)\phi_\downarrow(x)
+\phi_\downarrow(x,\tau)\phi_\uparrow(x) \rangle $, where $\tau$
is the imaginary time. When the magnetization is not
unidirectional, on the other hand, other triplet components with
$S_z=\pm 1$ given by
$\langle\phi_\uparrow(x,\tau)\phi_\uparrow(x)\rangle$ and
$\langle\phi_\downarrow(x,\tau)\phi_\downarrow(x)\rangle$ are also
induced because $S_z=0$ along a spin quantization axis is not pure
$S_z=0$ but a mixture of $S_z=0$, $+1$, and $-1$ along a different
axis. The $S_z=\pm 1$ TPC has the long coherence length $\xi_F$
in F because the exchange field is not pair breaking for them.

We here note the following distinctive features of the TPC which
appears in S/F systems.\cite{Bergeret05rmp} See below for more
detailed discussion. (a) The S in an S/F bilayer is a conventional
$s$-wave singlet superconductor. The exotic state of OFTPC is
induced not because the S of the S/F is exotic. It is induced
because of the proximity effects and the broken time reversal
symmetry. (b) TPC is induced in addition to the dominant SPC, and
may have short coherence length $\xi_{ex}$ or long coherence length
$\xi_F$. (c) TPC is even in momentum but odd in frequency. (d) The
$T_c$ of an S/F bilayer is determined by the SPC (See Eq.\
(\ref{self-consistency}) below.), and TPC can change $T_c$
indirectly only by changing SPC through the boundary conditions. See
Sec.\ II in the following for detailed discussions. The OFTP is an
exotic state of matter, but it has not been directly observed yet. A
lot of experimental and theoretical work has focused on finding
fingerprints of the OFTP
correlations.\cite{Klapwijk06nature,Eschrig05prl} The basic idea is
that one convert the short range triplet component into the long
range one by controlling or inducing nonuniformity of the F
magnetization direction, and observe the long length scale triplet
component with appropriate experiments. Even when the exchange field
is uniform in the F, the long range components are induced when the
interface is spin active which causes spin flip or spin
rotation.\cite{Millis88prb,Sauls04prb,Belzig05prb,Tokuyasu88prb} The
present formulation can be straightforwardly applied in this case
too with a modified version of the boundary conditions (BC) given by
Eq.\ (\ref{eqn:bc_tri4}). Recently, there was a report which
strongly hints the existence of long length OFTPC in
S/F/S.\cite{Klapwijk06nature}

The superconductor-ferromagnet systems, as described above,
exhibit very interesting and exotic behavior, and it will be
important to understand them in a systematic way. To do that, we
have theoretically studied the S/N/F trilayer systems. By varying
the intervening N thickness $d_N$ and the interface resistances
between S and N and N and F, represented in terms of
$\gamma_b^{SN}$ and $\gamma_b^{NF}$, respectively, we can control
the proximity effects in a systematic way. In this paper, the
proximity effects in S/N/F trilayers are studied for $T_c$, SPC,
and TPC. This paper is organized as follows: After this section,
we present in Sec.\ II the generalization to S/N/F trilayers of
the numerically exact Green's function method developed by Fominov
$et~al.$\cite{Fominov02prb} for S/F bilayers. The resulting
Usadel equations subject to appropriate boundary conditions are
then solved self-consistently by numerical iterations for $T_c$,
SPC, and TPC. The results will be presented in Sec.\ III for
typical values of $\gamma_b^{SN}$, $\gamma_b^{NF}$, and $E_{ex}$.
Based on these results, we will make some comments on the
experimental observation of the odd frequency triplet components
and the recent $T_c$ measurements in Nb/Au/CoFe trilayer
systems.\cite{Char05prb} We then conclude with the summary and
concluding remarks in Sec.\ IV. In Appendix, we collect some
technical details for calculating the singlet and triplet pairing
components.

\section{Usadel formulation}
\label{sec:theory}

\subsection{Usadel equation for S/N/F trilayers}

\begin{figure}[hbt]
\epsfxsize=8cm \epsffile{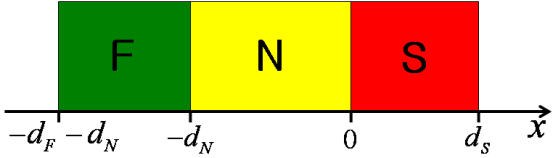} \caption{Schematic drawing of an
S/N/F trilayer. The coordinate $x$ is taken as perpendicular to
the interfaces. }\label{fig:snf}
\end{figure}

We wish to understand the proximity effects in the S/N/F trilayer
systems, which is schematically shown in Fig.\ \ref{fig:snf}. The S
is a conventional $s$-wave singlet pairing superconductor like Nb, N
is Au, Cu, or Al (above Al's $T_c$), and F is CoFe or Ni. Each of S,
N, F is characterized by the coherence length $\xi$ and resistivity
$\rho$. In addition, S layer is described by superconducting
critical temperature $T_{c0}$, and F by the exchange energy
$E_{ex}$. The layers of thin films are in the dirty limit and we
employ the dirty limit quasi-classical theory, the Usadel equation,
to describe them.\cite{Usadel70prl} Moreover, since we are
interested in calculating the superconducting transition temperature
$T_c$ and the pair functions near $T\approx T_c$ of S/N/F, we will
use the linearized Usadel equation. Near $T_c$, The normal Green
function is $G=sgn(\w_n) \delta_{\sigma,\sigma'}$, where $\sigma$
and $\sigma'$ are the spins of two electrons forming a Cooper pair,
and the anomalous function is $F\rightarrow 0$. The Usadel equation
is therefore linearized with respect to $F$. In a general case where
all four components of $F$ are kept, $F$ may be represented as a
$2\times2$ matrix or a 4 component vector. In the present problem
where the exchange field is uniform over the $F$ region, only two
components, SPC and short range TPC, appear in the equation and the
anomalous function $F$ may be represented as a two component vector
or scalar complex function. Because we will consider long range TPC
in a subsequent study where a vector representation is convenient we
will formulate the present problem in terms of two component vector
 \beqa
{\bf F}_i (x,i\w_n) = \left(\begin{array}{c} F_{i}^+(x,i\w_n)
\\F_{i}^- (x,i\w_n) \end{array} \right),
 \label{state vector}
 \eeqa
where $i=$ S, N, or F, and $x$ is the coordinate perpendicular to
the interface and the translational symmetry is assumed parallel
to the interface. The superscript $\pm$ refers to the even and odd
functions of the frequency.
 \beq
F^{\pm}(x,i\w_n) \equiv \f{F(x,i\w_n) \pm F(x,-i\w_n)}{2},
\label{eqn:trick}
 \eeq
where the anomalous function $F$ is defined by
 \beq
F(x,i\w_n)=-\int_0^\beta d\tau e^{i\w_n\tau} T_\tau\langle
\phi_\uparrow(x,\tau)\phi_\downarrow(x,0)\rangle,
 \eeq
where $T_\tau$ is the $\tau$ ordering operator. Then,
$F(x,-i\w_n)=\int_0^\beta d\tau e^{i\w_n\tau} T_\tau\langle
\phi_\downarrow(x,\tau)\phi_\uparrow(x,0)\rangle$, and the even
function $F^+$ represents the SPC and the odd function $F^-$ the
TPC of the anomalous function. We will also use the SPC (TPC) to
stand for $\Psi^+(x)$ ($\Psi^-(x)$) which is just the summation
over Matsubara frequencies of $F^+(x,i\w_n)$ ($F^-(x,i\w_n)$) as
given in Eq.\ (\ref{eqn:pair_amp1}).

The linearized Usadel equation near $T\approx T_c$ takes the
following form:\cite{Fominov02prb}
 \beqa
\label{eqn:Usadel-S} \xi_S^2\pi T_{c}\f{\partial^2}{\partial
x^2}{\bf F}_S(x,i\w_n)= |\w_n|{\bf F}_S(x,i\w_n)-{\bf
\D}(x),\nonumber\\ (0<x<d_S),
\\ \label{eqn:Usadel-N} \xi_N^2\pi T_{c}\f{\partial^2}{\partial
x^2}{\bf F}_N(x,i\w_n)= |\w_n|{\bf F}_N(x,i\w_n), \nonumber \\
(-d_N<x<0),\\ \label{eqn:Usadel-F} \xi_F^2\pi
{T_c}\f{\partial^2}{\partial x^2}{\bf F}_F(x,i\w_n)= |\w_n|{\bf
F}_F(x,i\w_n) \nonumber \\ +i\ \sgn(\w_n)\ E_\subbox{ex}\sigma_1
{\bf F}_F(x,i\w_n), \nonumber\\ (-d_F-d_N<x<-d_N).
 \eeqa
where $\w_n = 2\pi T(n+\f{1}{2})$ is the Matubara frequency and
$\sigma_1$ is the Pauli spin matrix. ${\bf \Delta}(x) =
\left(\begin{array}{c} \Delta(x)
\\0 \end{array} \right)$. Eq.\ (\ref{eqn:Usadel-F}) explicitly
shows that the exchange field $E_{ex}$ couples the SPC $F_{F}^+$
and TPC $F_{F}^-$ and induces TPC from SPC. The coherence length
is $\xi = \sqrt{\f{D}{2\pi T_{c}}}$ in the dirty limit where $D$
is the diffusion constant. Then,
 \beqa
\xi_S = \sqrt{\f{D_S}{2\pi T_{c}}},~~ \xi_N = \sqrt{\f{D_N}{2\pi
T_{c}}}, ~~\xi_F = \sqrt{\f{D_F}{2\pi T_{c}}}
 \eeqa
in each S, N, and F layer, as discussed in Introduction. The
singlet component $F_{S}^+$ determines the gap function
$\Delta(x)$
 \beq
\label{eqn:orderp} \D(x) = V g(\e_F) \pi T_c
\sum_{\w_n>0}F_{S}^{+}(x,i\w_n),
 \eeq
where $V$ and $g(\e_F)$ are, respectively, the pairing
interaction and the density of states per spin at the Fermi level.
It satisfies the self-consistency relation
 \beqa\label{self-consistency}
\Delta(x) \ln\left(\frac{T_{c0}}{T_c}\right) = \pi T_c
\sum_{\omega_n}
\left[\frac{\Delta(x)}{|\omega_n|}-F_{S}^+(x,i\w_n)\right],
 \eeqa
where $T_c$ is the superconducting critical temperature of the
S/N/F trilayer and $T_{c0}$ is that of the S layer alone without
the N and F layers. Eq.\ (\ref{self-consistency}) explicitly
shows that $T_c$ is determined by the singlet component alone.
The triplet components may affect $T_c$ only by changing the
singlet component through the boundary conditions discussed below.

Being a differential equation, the Usadel equation must be
supplemented by boundary conditions. The appropriate BC for the
Usadel equation are well established and given as
follows.\cite{Kupriyanov88jetp}
 \beqa \label{eqn:bc_tri1}
&&\f{d}{dx}{\bf F}_F(-d_F-d_N) = 0,\\
\label{eqn:bc_tri2}
&&\xi_N\f{d}{dx}{\bf F}_N(-d_N) - \g_{NF}\xi_F\f{d}{dx}{\bf F}_F(-d_N)= 0,\\
\label{eqn:bc_tri3}
&&\xi_S\f{d}{dx}{\bf F}_S(0) - \g_{SN}\xi_N\f{d}{dx}{\bf F}_N(0)= 0,\\
\label{eqn:bc_tri4} &&{\bf F}_N(-d_N)-{\bf F}_F(-d_N)
=\g_b^{NF}\xi_F\f{d}{dx}{\bf F}_F(-d_N), \\
\label{eqn:bc_tri5} &&{\bf F}_S(0)-{\bf F}_N(0)=
\g_b^{SN}\xi_N\f{d}{dx} {\bf F}_N(0), \\
\label{eqn:bc_tri6} &&\f{d}{d x}{\bf F}_S(d_S) =0,
 \eeqa
where
 \beq
\g_{NF}\equiv \f{\rho_N\xi_N}{\rho_F\xi_F}, ~~\g_{SN}\equiv
\f{\rho_S\xi_S}{\rho_N\xi_N}.
 \eeq
The $\g_b^{SN}$ and $\g_b^{NF}$ are dimensionless quantities
characterizing S/N and N/F interfaces, respectively, given by
 \beqa \g_{b}^{NF}\equiv
\f{R_b^{NF}\cA}{\rho_F\xi_F},\hspace{0.5cm} \g_{b}^{SN}\equiv
\f{R_b^{SN}\cA}{\rho_N\xi_N},
 \eeqa
where $R_b^{NF}$ and $R_b^{SN}$ are the interface resistances
between N and F and S and N, respectively, and $\cA$ is the
interface area. $\gamma_b^{SN}$ and $\gamma_b^{NF}$ are modeled in
terms of spin conserving potential
barriers,\cite{Zaitsev84jetp,Kupriyanov88jetp} and do not introduce
any spin modifying mechanism. On the other hand, the boundary of the
F can be spin dependent\cite{Perez-Willard04prb} or spin
active,\cite{Millis88prb,Sauls04prb,Belzig05prb,Tokuyasu88prb} and
the BC of Eq.\ (\ref{eqn:bc_tri4}) should be modified accordingly.
The spin-active interface can induce the long range triplet
components. One possible description of the spin-active interface is
that the parameter $\gamma_b^{NF}$ is given by a $4\times 4$ matrix
instead of a scalar with the state of Eq.\ (\ref{state vector})
enlarged to a four component vector. All of the four components, two
of which are the long range ones, are non-vanishing in general
because they are coupled via the BC Eq.\ (\ref{eqn:bc_tri4}).

\subsection{Boundary condition in terms of ${\bf F}_S$}

The Usadel equation \eqn{eqn:Usadel-S}, \eqn{eqn:Usadel-N},
\eqn{eqn:Usadel-F} together with the BC \eqn{eqn:bc_tri1},
\eqn{eqn:bc_tri2}, \eqn{eqn:bc_tri3}, \eqn{eqn:bc_tri4},
\eqn{eqn:bc_tri5}, \eqn{eqn:bc_tri6} and self-consistency equation
\eqn{self-consistency} forms a complete set of equations to
describe an S/N/F trilayer near $T\approx T_c$. It is solved by
extending the numerically exact Green's function technique
developed by Fominov $et ~al.$ for the S/F
bilayers.\cite{Fominov02prb,Arfken01book} To utilize the
technique, everything should be written in terms of ${\bf F}_S$
alone in the S region. The Usadel equation and the
self-consistency relation are already written in terms of ${\bf
F}_S$ alone in the S region, and the remaining task is to write
the BC at $x=0$ and $x=d_S$ in terms of ${\bf F}_S$ only. The BC
at $x=d_S$ is already given by Eq.\ (\ref{eqn:bc_tri6}), and the
one at $x=0$ can be derived by writing the Usadel equation with
the corresponding BC successively starting from the F to N
regions. The procedure is very similar to Fominov
$et~al.$\cite{Fominov02prb}

First, for the F region we solve the homogeneous equation
\eqn{eqn:Usadel-F} with BC \eqn{eqn:bc_tri1} to obtain
  \beqa
\label{eqn:sol_tri_f} \bF_F =\left(\begin{array}{cc}\cosh
k_F(x_{NF}) & \cosh k_F^*(x_{NF}) \\ -\cosh k_F(x_{NF}) & \cosh
k_F^*(x_{NF})\end{array} \right)
 \left(\begin{array}{c}
C^+_{F}(i\w_n)\\C^-_{F}(i\w_n)\end{array}\right),
 \eeqa
where
 \beq
x_{NF}=x+d_N+d_F,~~ k_F=\f{1}{\xi_F}\sqrt{\f{|\w_n|+iE_{ex}}{\pi
T_{c}}} .
 \eeq
Note that the $k_F$ implies another length scale $\xi_{ex}$ much
shorter than $\xi_F$ alluded earlier. Because $E_{ex} \gg |\w_n|$
for relevant Matsubara frequencies, one has
 \beq
\xi_{ex}=\frac{1}{Re \left(k_F\right)}=\sqrt{\frac{D_F}{E_{ex}}}.
\label{xi_ex}
 \eeq
The length scale of pair function is given by $\xi_{ex}$ in F
region. From \eqn{eqn:sol_tri_f} we obtain
 \beqa
\xi_F\f{d}{dx}\bF_F(-d_N)=\hat{\cA}_F\bF_F(-d_N),\label{eqn:Af}
 \eeqa
where
 \beqa \hat{\cA}_F&\equiv& \left(\begin{array}{cc}
\re A_{F}& -i\im A_{F}  \\
-i\im A_{F}& \re A_{F} \end{array} \right), \\ A_F &\equiv&
k_F\xi_F\tanh k_Fd_F.
 \eeqa
Then, utilizing the BC at $x=-d_N$ of Eq.\ (\ref{eqn:bc_tri2}) and
(\ref{eqn:bc_tri4}), the ${\bf F}_N(-d_N)$ and $\frac{d}{dx}{\bf
F}_N(-d_N)$ can be written in terms of ${\bf F}_F(-d_N)$ and
$\frac{d}{dx}{\bf F}_F(-d_N)$. Then, using Eq.\ (\ref{eqn:Af}),
we find
 \beqa\label{eqn:An-dn}
\xi_N\f{d}{dx}\bF_N(-d_N)=\hat{\cA}_{NF}\bF_N(-d_N),
 \eeqa
where
 \beq \hat{\cA}_{NF} \equiv
\left(1+\hat{\cA}_F\g_b^{NF}\right)^{-1} {\g_{NF}\hat{\cA}_F} .
 \eeq

Second, the anomalous function ${\bf F}_N$ in the N region
described by Eq.\ \eqn{eqn:Usadel-N} may be written as
 \beqa \label{eqn:sol_tri2}
{\bf F}_N(x,i\w_n)={\bf F}_N(-d_N)\cosh k_N(x+d_N)\nonumber\\
+\xi_N\f{d}{dx}{\bf F}_N(-d_N) \f{\sinh k_N(x+d_N)}{k_N\xi_N},
 \eeqa
where
 \beq k_N=\f{1}{\xi_N}\sqrt{\f{|\w_n|}{\pi T_{c}}}.
 \eeq
Then, we use Eqs.\ (\ref{eqn:sol_tri2}) and (\ref{eqn:An-dn}) to
obtain
 \beqa
\xi_N\f{d}{dx}\bF_N(0)=\hat{\cA}_{NS}\bF_N(0),\label{eqn:ANS}
 \eeqa
where
 \beqa \hat{\cA}_{NS} \equiv
\left(A_N+\hat{\cA}_{NF}\right)
\left(1+\f{\hat{\cA}_{NF}A_N}{k_N^2\xi_N^2}\right)^{-1},\\
A_N = k_N\xi_N\tanh k_N d_N.
 \eeqa

Third, using BC at $x=0$ given by Eqs.\ (\ref{eqn:bc_tri3}) and
(\ref{eqn:bc_tri5}), and Eq.\ (\ref{eqn:ANS}), we can finally
obtain the BC in terms of ${\bf F}_S(0)$ and $\frac{d}{dx}{\bf
F}_S(0)$ such as
 \beqa
\xi_S\f{d\bF_S(0)}{dx}=\hat{A}_S\bF_S(0), \label{eqn:bc_tri_s}
 \eeqa
where
 \beq \hat{\cA}_S \equiv
\left(1+\hat{\cA}_{NS}\g_b^{SN}\right)^{-1} \g_{SN}\hat{\cA}_{NS}.
\label{eqn:AS}
 \eeq
Eq.\ (\ref{eqn:bc_tri_s}) connects $\f{dF_S^+(0)}{dx}$ and
$\f{dF_S^-(0)}{dx}$ with $F_S^+(0)$ and $F_S^-(0)$. But, we need
the BC which connect $\f{dF_S^+(0)}{dx}$ with $F_S^+(0)$ only to
solve the Usadel equation in S region for $0 < x < d_S$ given by
Eq.\ (\ref{eqn:Usadel-S}).
 \beq
\label{eqn:bi-usadel_ss} \pi T_c\xi_S^2\f{\partial^2}{\partial
x^2} F_{S}^+ (x,i\w_n) -|\w_n| F_{S}^+ (x,i\w_n) = - \D(x).
 \eeq
We write from BC of (\ref{eqn:bc_tri6})
 \beq\label{sol:fst}
F_S^-(x,i\w_n)=C_S^-(i\w_n) \cosh k_S(x-d_s),
 \eeq
where
 \beq
k_S = \frac{1}{\xi_S} \sqrt{\frac{|\w_n|}{\pi T_c}}.
 \eeq
Using this to eliminate the $F_S^-(0)$ from \eqn{eqn:bc_tri_s}
for $\f{dF_S^+(0)}{dx}$, we obtain
 \beqa
\xi_S\f{dF_S^+(0)}{dx}=W(i\w_n)F_S^+(0), \label{eqn:bi-S}
 \eeqa
where
 \beqa W(i\w_n)&=&\g_{SN}\f{A_S(\g_b^{SN}+ Re
B_{SN})+\g_{SN}}{A_S|\g_b^{SN}+B_{SN}|^2 +\g_{SN}(\g_b^{SN}+Re
B_{SN})},\nonumber\\
 \label{eqn:Wwn}
 \eeqa
and $A_S$ and $B_{SN}$ are defined by
 \beqa
A_S&=&k_S\xi_S\tanh k_Sd_S,\\
B_{SN}&=& \left[k_N\xi_N\tanh
k_N\left(d_N+x_0\right)\right]^{-1}\\
\tanh k_Nx_0&=& \frac{1}{k_N \xi_N}
\frac{\gamma_{NF}}{\gamma_b^{NF}+A_F^{-1}}.
 \eeqa

\subsection{Green's function method}

Now, the problem is reduced to solving Eq.\
(\ref{eqn:bi-usadel_ss}) with BC \eqn{eqn:bc_tri6} and
\eqn{eqn:bi-S}. To do it by the Green's function technique, one
needs to solve the following source
equation.\cite{Fominov02prb,Arfken01book}
 \beq \label{eqn:source} \pi
T_c\xi_S^2\f{\partial^2}{\partial x^2} G(x,y) -|\w_n|G(x,y) = -
\d(x-y),
 \eeq
with the BC corresponding to \eqn{eqn:bi-S} and \eqn{eqn:bc_tri6}
 \beqa
\xi_S\f{\partial}{\partial x}G(0,y)
&=&W(i\w_n)G(0,y),\\
\xi_S\f{\partial}{\partial x}G(d_S,y)&=&0.
 \eeqa
The Green's function can be constructed as
follows.\cite{Arfken01book}
 \beqa
\nonumber G(x,y;i\w_n)&=& \f{k_S/|\w_n|}
{\sinh k_Sd_S +\f{W(i\w_n)}{k_S\xi_S}\cosh k_S d_S}\\
&&\times\left\{
\begin{array}{cc}
v_1(x)v_2(y),& 0<x<y \\
v_1(y)v_2(x),& y<x<d_S
\end{array}
\right. ,
 \eeqa
where
 \beqa v_1(x)&=&\cosh k_S x +\f{W(i\w_n)}{k_S\xi_S}\sinh k_S x , \\
v_2(x)&=&\cosh k_S(x-d_S).
 \eeqa
The solution of \eqn{eqn:bi-usadel_ss} is then
 \beq
\label{eqn:bi-ss-sol} F_{S}^{+}(x,i\w_n) = \int_0^{d_S}dy
G(x,y;i\w_n)\D(y).
 \eeq
The gap function $\D(x)$ is determined by Eq.\ (\ref{eqn:orderp}).

Substituting \eqn{eqn:bi-ss-sol} into \eqn{eqn:orderp} gives the
self-consistency equation,
 \beq \D(x) = V
g(\e_F) \pi T_c \sum_{\w_n>0}\int_0^{d_S}dy G(x,y;i\w_n)\D(y).
 \eeq
From this self-consistency equation, we can write down the
equation for $T_c$ with respect to $T_{c0}$.
 \beqa\label{eqn:Self_TC}
\D(x) \ln\left(\f{T_{c0}}{T_{c}}\right)=
\int_0^{d_S} dy~M(x,y) \D(y) , \nonumber \\
M(x,y) \equiv \pi T_c \sum_{\w_n>0}\left[ \f{\delta(x-y)}{|\w_n|}
-G(x,y;i\w_n)\right].
 \eeqa
The above integral equation can be reduced to an eigenvalue
problem after we change the integration over $y$ into a summation
by discretization. The $T_c$ is determined by the smallest
eigenvalue of the discretized $M(x,y)$ matrix, which yields the
highest $T_c$. One plugs an arbitrary initial $T_c$ into the
right hand side of Eq.\ (\ref{eqn:Self_TC}) and calculates a new
$T_c$, and iterates until the input and output are equal within a
tolerance. After a convergence is reached, $T_c$ is obtained, and
the corresponding eigenvector to the smallest eigenvalue gives
$\Delta(x)$. Then, the anomalous function $F^+$ can be obtained
from Eq.\ (\ref{eqn:bi-ss-sol}), and the singlet and triplet pair
functions $\Psi^+(x)$ and $\Psi^-(x)$ can be calculated from
 \beq \label{eqn:pair_amp1}\Psi^\pm (x) = \pi
T_c\sum_{\w_n>0}F^{\pm}(x,i\w_n).
 \eeq
Technical details for calculating SPC and TPC are collected in
Appendix.

\section{Results: $T_c$ and pair functions of S/N/F trilayers}
\label{sec:result}

\subsection{Perfect interfaces}
\label{sec:singlet}

$T_c$ of S/N/F trilayers were calculated by self-consistently
solving Eq.\ (\ref{eqn:Self_TC}) by numerical iterations. And
$T_c$ of S/N bilayers were calculated for comparison. The singlet
and triplet pair functions were also calculated as detailed in
Appendix. Let us first consider the ideal case of perfect
interfaces with $\gamma_b^{SN} = \gamma_b^{NF} =0$. We take
$\rho_S = 15.9627$ $\m\W$cm, $\rho_N = 2.0$ $\m\W$cm, $\rho_F =
40.0$ $\m\W$cm, $\xi_S = 7.0$ nm, $\xi_N = 110$ nm, $\xi_F =
10.241$ nm, $E_{ex} = 1235$ K, and $T_{c0}=7.927$ K, which are
appropriate for S = 23 nm Nb, N = Au, and F = CoFe as reported in
the reference \cite{Char05prb,Char05prbSF}. Recall that the
condition $E_{ex}\tau \ll 1$ should be satisfied for the
diffusive Usadel equation to be applicable. Using $D=\frac13
v_F^2 \tau$, and the $\xi_F$ and $T_{c0}$ values, we obtain
$E_{ex}\tau \approx 10^{-2}$ so that the Usadel equation is
applicable to Nb/CoFe systems.

\begin{figure}[hbt]
\epsfxsize=8cm \epsffile{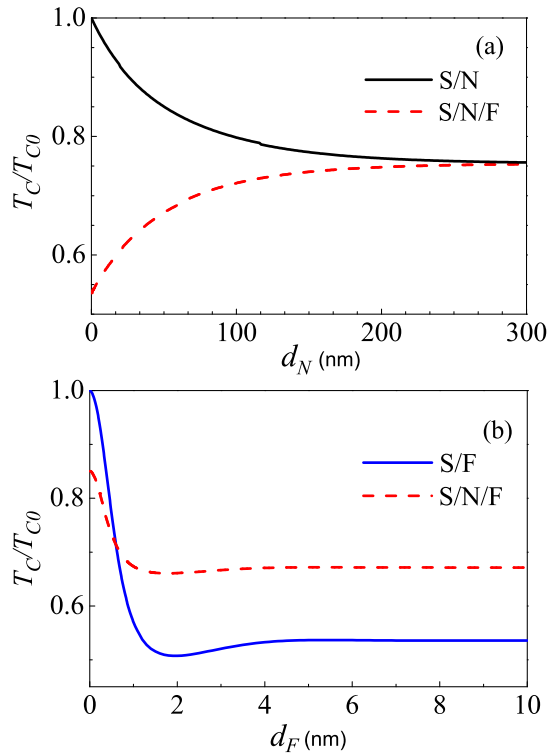} \caption{ (a)
Numerical results of $T_c/T_{c0}$ of S/N/F and S/N as a function
of $d_N$, shown by dashed and solid curves, respectively. Both
curves have the coherence length $\xi_N$ and merge as
$d_N\rightarrow\infty$ as expected. (b) $T_c/T_{c0}$ of S/N/F and
S/F as a function of $d_F$, shown by the dashed and solid curves.
Note the different length scales $\xi_{N}$ and $\xi_{ex}$ between
(a) and (b). }\label{fig:SN-SNF-SF-Tc}
\end{figure}

We show in Fig.\ \ref{fig:SN-SNF-SF-Tc}(a) $T_c$ as a function of
N layer thickness, $d_N$, for $d_S=23$ nm and $d_F=23$ nm S/N and
S/N/F. For S/N bilayers $T_c(d_N)$ decreases as $d_N$ is
increased, but for S/N/F trilayers $T_c(d_N)$ increases as $d_N$
is increased. The length scale of both decrease and increase is
given by the N coherence length $\xi_N$. As $d_N\rightarrow
\infty$ the two $T_c$ approach each other as expected. In Fig.\
\ref{fig:SN-SNF-SF-Tc}(b) we show $T_c$ as a function of $d_F$
for S/F and S/N/F. Note the different length scales between (a)
and (b) given by $\xi_N$ and $\xi_{ex}$, respectively. $T_c(d_F)$
for S/F shows the well known behavior of an exponential decrease
with a shallow dip. For S/N/F, we took $d_N=50$ nm. The presence
of the N layer weakens the effects of exchange energy of F.
Consequently, $T_c(d_F)$ for S/N/F, compared with S/F, is lower
at $d_F=0$ but higher as $d_F \rightarrow \infty$, and has a
shallower dip.

\begin{figure}[hbt]
\epsfxsize=9cm \epsffile{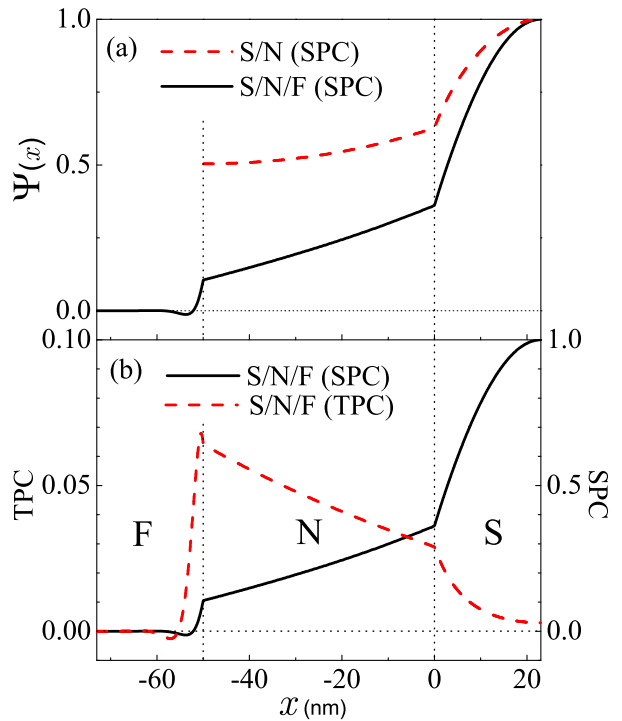} \caption{(a) SPC of S/N/F
and S/N shown by the solid and dashed lines, respectively.
$\Psi(x)$ is normalized by the value at the outer surface of the
S layer. (b) SPC and TPC of an S/N/F trilayer shown by solid and
dashed lines, respectively. TPC is magnified by 10 times for
clarity. }\label{fig:sn-snf-PA}
\end{figure}

We may also calculate the pair functions as well as $T_c$ from
the Usadel equation for various S-N-F layers. In Fig.\
\ref{fig:sn-snf-PA}(a) we show the SPC $\Psi^+(x)$ for S/N and
S/N/F for $d_S=23$ nm, $d_N=50$ nm, and $d_F=23$ nm, normalized
by the value at $x=23$ nm. It decreases monotonically in N but
non-monotonically in F. The non-monotonicity of $\Psi^+(x)$ and
$T_c(d_F)$ results from the same physics. It may be understood as
originating from the non-zero net momentum of a Cooper
pair.\cite{Demler97prb} In (b) we show both SPC and TPC for an
S/N/F. While SPC is maximum at the S outer surface of $x=23$ nm,
TPC is maximum near the N/F interface. The position of TPC peak is
roughly $\xi_{ex}$ from the N/F interface.

\begin{figure}[hbt]
\epsfxsize=9cm \epsffile{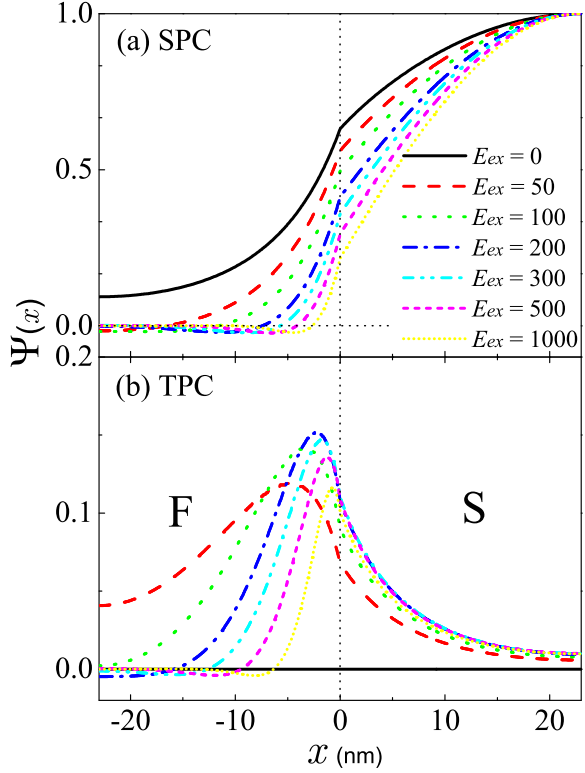} \caption{SPC and TPC of S/F
bilayers for several $E_{ex}$. The peak position of TPC is given
by the decay length scale $\xi_{ex}$ of SPC. Note that the TPC
vanishes as F becomes an N ($E_{ex}=0$. }\label{fig:sf-PA}
\end{figure}

To see the TPC more clearly, we plot in Fig.\ \ref{fig:sf-PA} the
SPC and TCP of S/F bilayers for several values of $E_{ex}$. Note
that for $E_{ex}=0$ F becomes like an N; the length scale
$\xi_{ex}$ becomes $\xi_F$ and TPC vanishes. The TPC becomes small
also for large $E_{ex}$ because TPC decays rapidly for small
$\xi_{ex}\propto E_{ex}^{-1/2}$. The TPC $\rightarrow 0$ as
$E_{ex}\rightarrow 0$ or $\rightarrow\infty$, and is fully induced
for an intermediate $E_{ex}$, as can be seen from figure (b). For
each $E_{ex}$, the peak of TPC is determined by the competition
between the exponential decrease of length scale $\xi_{ex}$ and
linear increase determined by the BC
 \beq
\xi_S\f{d}{dx}{\bf F}_S(0) - \g_{SF}\xi_F\f{d}{dx}{\bf F}_F(0)= 0.
 \eeq
The TPC, therefore, has a peak at around $\xi_{ex}$ away from S/F
interface, which can be seen from Fig.\ \ref{fig:sf-PA}.

\subsection{Effects of interface resistances}

\begin{figure}[hbt]
\epsfxsize=9cm \epsffile{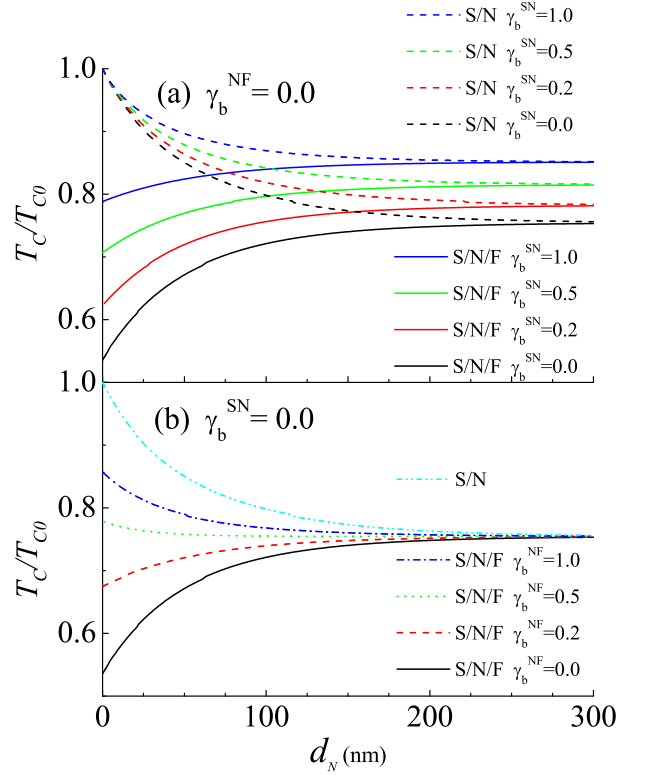} \caption{$T_c(d_N)$ of
S/N/F and S/N for several different interface resistances. (a) is
for several $\gamma_b^{SN}$ with $\gamma_b^{NF}=0$. Two $T_c(d_N)$
curves of S/N/F and S/N merge as $d_N\rightarrow \infty$. (b) is
for several $\gamma_b^{NF}$ with $\gamma_b^{SN}=0$. Note that for
small $\gamma_b^{NF}$ $T_c$ increases as $d_N$ is increased but
decreases for large $\gamma_b^{NF}$. } \label{fig:gamma-Tc}
\end{figure}

We now turn to the more realistic case with non-zero interface
resistances. There are two interfaces in an S/N/F, between S and
N, and N and F, parameterized by $\gamma_b^{SN}$ and
$\gamma_b^{NF}$, respectively. The main effect of the interface
resistances is to weaken the proximity effects and increase the
$T_c$ of S/N/F trilayers. This can be seen from Fig.\
\ref{fig:gamma-Tc}. Figure (a) shows $T_c$ vs.\ $d_N$ of S/N/F for
several $\gamma_b^{SN}$ with $\gamma_b^{NF}=0$. $T_c$ vs.\ $d_N$
of S/N with the same $\gamma_b^{SN}$ are also shown for
comparison. Figure (b) shows $T_c$ vs.\ $d_N$ for several
$\gamma_b^{NF}$ with $\gamma_b^{SN}=0$. In both figures (a) and
(b), $T_c$ of a given $d_N$ increases as $\gamma_b$ is increased.

\begin{figure}[hbt]
\epsfxsize=9cm \epsffile{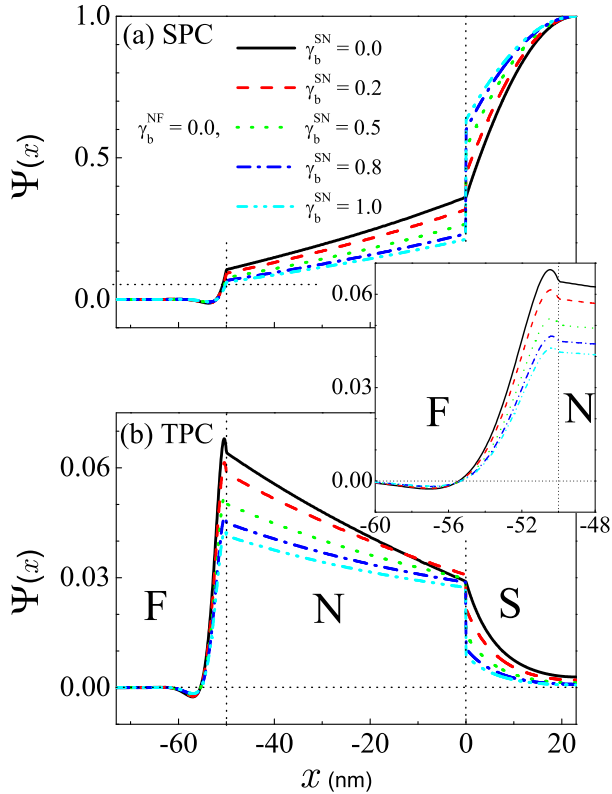} \caption{The SPC and TPC of
S/N/F for several $\gamma_b^{SN}$ with $\g_b^{NF}=0$. The inset
shows TPC in detail in F region. Non-zero $\gamma_b^{SN}$ produces
a discontinuity in SPC and TPC at the S/N interface.
}\label{fig:gbSN-PA}
\end{figure}

\begin{figure}[hbt]
\epsfxsize=9cm \epsffile{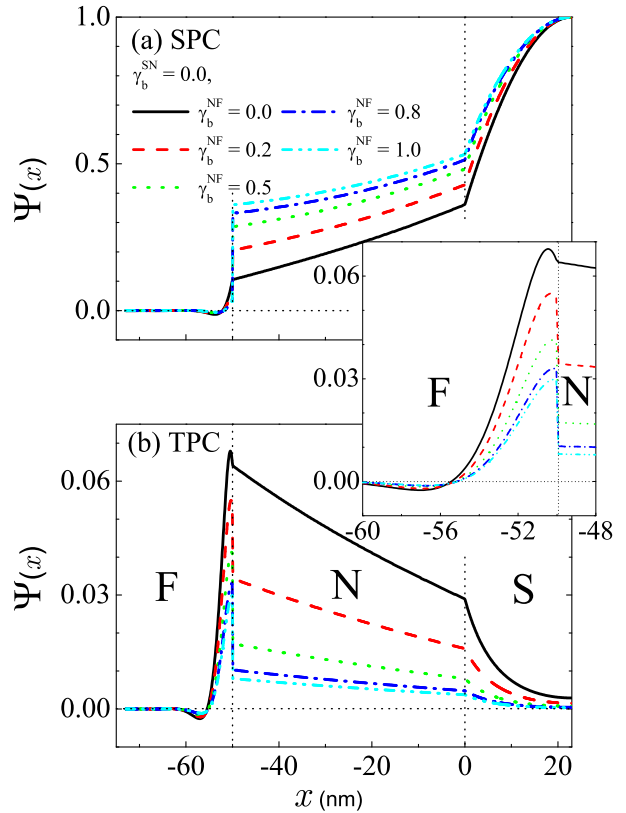} \caption{The SPC and TPC of
S/N/F for several $\gamma_b^{NF}$ with $\g_b^{SN}=0$. The inset
shows the TPC in detail in F region. Non-zero $\gamma_b^{NF}$
introduces the discontinuity at the N/F interface. The TPC in F
region is suppressed by $\gamma_b^{NF}$ less than SPC in F region
or TPC in N region. }\label{fig:gbNF-PA}
\end{figure}

Another, perhaps more technical, way of saying this is from BC of
Eqs.\ (\ref{eqn:bc_tri2}), (\ref{eqn:bc_tri3}),
(\ref{eqn:bc_tri4}), and (\ref{eqn:bc_tri5}). Non-zero
$\gamma_b^{SN}$ ($\gamma_b^{NF}$) introduces the discontinuity of
the anomalous function $F^\pm$ at the interface. The anomalous
function $F^+$ and pair function $\Psi^+$, therefore, increases in
S but decreases in F region as shown in Figs.\ \ref{fig:gbSN-PA}
and \ref{fig:gbNF-PA}. An increase of $\Psi^+(x)$ in S means an
increase of $T_c$ because it is the largest value of $\Psi^+(x)$
that determines $T_c$. (In the figures, however, $\Psi^+(x)$ is
normalized at $x=d_S$ and this point is not manifested clearly.)
We plot in Fig.\ \ref{fig:gbSN-PA} $\Psi^\pm(x)$ for S/N/F for
several $\gamma_b^{SN}$ values with $\g_b^{NF}=0$. And in Fig.\
\ref{fig:gbNF-PA} $\Psi^\pm(x)$ for several $\gamma_b^{NF}$
values with $\g_b^{SN}=0$. Note that $\Psi^+(x)$ in S region is
increased as $\gamma_b^{SN}$ or $\gamma_b^{NF}$ is increased.

Now, let us look at in more detail how $T_c$ behaves as a function
of $d_N$ for non-zero $\gamma_b^{SN}$ or $\gamma_b^{NF}$. When
$\gamma_b^{NF}=0$ and $\gamma_b^{SN}$ is increased for S/N/F, as
shown in Fig.\ \ref{fig:gamma-Tc}(a), $T_c$ always increases as
$d_N$ is increased. However, when $\gamma_b^{NF}$ is increased
with $\gamma_b^{SN}=0$, $T_c(d_N)$ as a function of $d_N$ behaves
somewhat differently. For instance, let us compare the two
$T_c(d_N)$ curves for $\g_b^{NF}=0$ and $\g_b^{NF}= 1.0$.
$T_c(d_N)$ increases with $d_N$ for $\g_b^{NF}=0$ but decreases
for $\g_b^{NF}= 1.0$, and the two curves merge together as
$d_N\rightarrow\infty$. This is simple to understand. When
$\g_b^{NF}=0.0$, the pair breaking effects of the exchange field
of F fully influences S for small $d_N$, and its effect is
weakened as $d_N$ is increased. $T_c$, therefore, increases as
$d_N$ is increased. When $\g_b^{NF}=1.0$, on the other hand, the
effects of F is almost blocked out and it behaves like S/N.
Consequently, $T_c$ of $\g_b^{NF}=1.0$ is much larger compared
with that of $\g_b^{NF}=0.0$ as $d_N\rightarrow 0$ and decreases
as $d_N$ is increased. When $d_N \gtrsim$ 200 nm, the effects of
F are also blocked out irrespective of $\gamma_b^{NF}$, and all
$T_c$ curves merge together.

We now turn to the calculations of pair functions for S/N/F. In
Fig.\ \ref{fig:gbSN-PA} we took $\gamma_b^{NF}=0$ and plot SPC
and TPC for several $\gamma_b^{SN}$. Non-zero $\gamma_b^{SN}$
introduces the discontinuity in both SPC and TPC. The inset shows
the peak and dip structure of TPC more closely. As discussed for
Fig.\ \ref{fig:sf-PA}, the TPC has a peak around $\xi_{ex}$ away
from the N/F interface. In Fig.\ \ref{fig:gbNF-PA} we plot SPC
and TPC for several $\gamma_b^{NF}$ with $\gamma_b^{SN}=0$. The
discontinuity is at the N/F interface in this case.

\subsection{$T_c$ discontinuity due to boundary condition mismatch}

\begin{figure}[hbt]
\epsfxsize=8cm \epsffile{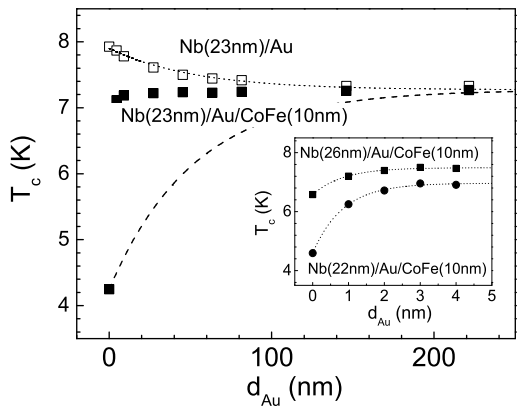} \caption{$T_c(d_N)$ of
Nb/Au/CoFe trilayers and Nb/Au bilayers, represented by the solid
and empty squares, respectively. Nb and CoFe layers are fixed at
$d_S=23$ and $d_F=10$ nm. The inset is the detailed plot of the
short length scale. The long dashed curve in the main plot is the
expected behavior of $T_c(d_N)$ without the BC mismatch. The
dotted line connecting empty squares is a guide to eyes.
Reproduced from the reference \onlinecite{Char05prb}.
}\label{fig:tcAu}
\end{figure}

An interesting point for $T_c(d_N)$ of S/N/F trilayers appears
when we consider $d_N\rightarrow 0$ limit. $T_c$ of this limit,
of course, ought to coincide with that of the corresponding S/F
of $d_N=0$. For example, $T_c$ of Nb/Au/CoFe trilayers must
approach that of Nb/CoFe as Au thickness goes to 0. Also, the
change of $T_c$ with $d_N$ should have the length scale of $\xi_N$
simply because there is no other length scale in the N region. We
therefore expect that $T_c(d_N)$ of Nb/Au/CoFe trilayers should
behave like the curve shown by the long dashed line beginning with
the $T_c$ of Nb/CoFe represented by the solid square at the low
left corner of the main plot in Fig.\ \ref{fig:tcAu}. The
experimental $T_c(d_N)$ measurements, however, show quite
different behavior as shown by the solid squares reproduced in
Fig.\ \ref{fig:tcAu}.\cite{Char05prb} There is an abrupt increase
of $T_c$ with the length scale of approximately 2 nm as $d_N$ is
increased as shown in the inset, and $T_c$ increases with the
expected length scale $\xi_N$ as $d_N$ is further increased.

One may understand this experimental $T_c$ vs.\ $d_N$ behavior as
follows. Recall that in the present formalism the interfaces are
modeled in terms of the two parameters $\gamma_b^{SN}$ and
$\gamma_b^{NF}$ irrespective of N thickness. S/N/F trilayers, as
$d_{N}\rightarrow 0$, still have two interfaces characterized by
$\gamma_b^{SN}$ and $\gamma_b^{NF}$, but the corresponding S/F
bilayer has one interface characterized by $\gamma_b^{SF}$.
Unless the three $\gamma_b$s satisfy a special match condition,
$T_c(d_N=0)$ of the bilayer and $\lim_{d_N\rightarrow 0}T_c(d_N)$
of the trilayers need not be the same. This condition may be
derived from the BC of Eqs.\ (\ref{eqn:bc_tri2}),
(\ref{eqn:bc_tri4}), and (\ref{eqn:bc_tri5}). We add
\eqn{eqn:bc_tri4} and \eqn{eqn:bc_tri5}, and take $d_N\rightarrow
0$. We then obtain using \eqn{eqn:bc_tri2}
 \beq
{\bf F}_S(0)-{\bf F}_F
(0)=\left(\gamma_b^{NF}+\gamma_b^{SN}\gamma_{NF}\right)\xi_F\frac{d}{dx}{\bf
F}_F(0).
 \eeq
For this to agree with the corresponding S/F bilayer, it is
required that
 \beqa
\gamma_b^{SF}=\gamma_b^{NF}+\gamma_b^{SN}\gamma_{NF}.
\label{match}
 \eeqa
This condition need not be satisfied. The $\gamma_b^{SF}$ may
take an arbitrary value irrespective of $\gamma_b^{NF}$ or
$\gamma_b^{SN}$. In case of a mismatch, $T_c(d_N)$ then shows a
discontinuity at $d_N=0$. The right hand side is usually larger
as is the case with the Nb/Au/CoFe trilayers, and $T_c$ of
$d_N\rightarrow 0$ S/N/F should be larger than that of S/F.

Note, however, that in experiments $d_N\rightarrow 0$ has a single
interface of course, and $T_c$ of Nb/Au/CoFe as $d_N\rightarrow
0$ must agree with that of Nb/CoFe. As $d_N$ is increased from 0
in Nb/Au/CoFe, there begin to form two distinct interfaces. For
the electrons to feel the two interfaces as the theoretical model
assumed, a finite width of N region is necessary. This width of
about 2 nm is the length which is needed to interpolate the $T_c$
of Nb/CoFe and that of $d_N\rightarrow 0$ Nb/Au/CoFe. The
theoretical step function behavior of $T_c(d_N)$ then should
appear as a short length scale of $\approx 2$ nm observed in the
experiments.\cite{Char05prb} This length scale is not the
material width of the Nb/Au interface which is approximately an
order of magnitude smaller. The same short length of $\approx2$
nm has also been observed for N = Cu,\cite{Char06Cu}
Al.\cite{Char06Al} The sudden increase of $T_c(d_N)$ as $d_N$ is
increased was also observed in the epitaxial
Nb(110)/Au(111)/Fe(110) trilayers by Yamazaki and his
coworkers.\cite{Yamazaki06prb}

\section{Summary and concluding remarks}
\label{sec:conclusion}

In this paper, we have extended the Green's function method
developed for S/F bilayers by Fominov $et~al.$ to S/N/F trilayer
systems. We have calculated the superconducting transition
temperature $T_c$ and the singlet and triplet pairing components
near $T\approx T_c$ for S/N/F trilayers. The interface between S and
N (N and F) was modeled in terms of $\gamma_b^{SN}$
($\gamma_b^{NF}$). They represent spin independent interface
resistances. As have been and will be reported separately, the
experimental $T_c$ measurements could be fit by the present
formalism.\cite{Char05prb,Char05prbSF,Char06Cu,Char06Al}

There seems, however, a room to improve the present formalism. One
possible route might be to model the N/F interface with two spin
dependent parameters as was done for example in ref.\
\onlinecite{Perez-Willard04prb}, or to model it in terms of a
spin-active
interface.\cite{Millis88prb,Sauls04prb,Belzig05prb,Tokuyasu88prb}
The latter route will be an interesting problem in connection with
the long range odd frequency triplet pairing components because they
are induced even when the exchange field in F region is uniform if
the N/F interface in spin-active.

Also noteworthy in the S/N/F system is the observation of the
length scale of $\approx 20$ nm in the $T_c$ vs. $d_N$
measurements in Nb/Au/CoFe trilayers.\cite{Char05prb} The natural
length scale in N region is the coherence length $\xi_N\approx
100$ nm as shown, for instance, in Figs.\ \ref{fig:gamma-Tc} and
\ref{fig:tcAu}. In addition, the short length of $\approx 2$ nm
was observed because of the boundary condition mismatch. In the
same experiments the $T_c$ as a function of $d_N$ also exhibited
a small oscillation with the length scale of $\approx 20$ nm. The
origin of this is not clear. It seems difficult to understand this
oscillation within the current Usadel formalism because there is
only one length scale of $\xi_N$ in the theory. It is interesting
that this $T_c$ oscillation was also observed in the epitaxial
Nb(110)/Au(111)/Fe(110) trilayers.\cite{Yamazaki06prb} In the
epitaxially grown Nb(110)/Au(111)/Fe(110) trilayers the $T_c$
oscillation length scale was $\approx 2$ nm, which is an order of
magnitude smaller than that of the sputtering grown Nb/Au/CoFe
trilayers. More experiments with different conditions will be
necessary to clarify the origin of the $T_c$ oscillations as a
function of $d_N$.

We have calculated the singlet pairing component and short range odd
frequency triplet pairing components in the present paper. The
triplet pairing component is induced on top of the dominant singlet
component by the proximity effects and time reversal symmetry
breaking exchange field. In the present work, the exchange field is
unidirectional over the F region, and only the short range triplet
component was induced. If the orientation of the exchange field
changes in the F region, or if the inner-interface of F is spin
active, then the other long range triplet components will also
appear. In the latter case the procedures developed in the present
manuscript can be directly applied to understand the long range odd
frequency triplet components. We are currently investigating the
spectroscopic properties of them in various trilayers of
superconductor and ferromagnet where the present formalism can be
applied.

\begin{acknowledgements}

This work was supported by Korea Science \& Engineering
Foundation (KOSEF) through Basic Research Program Grant No.\
R01-2006-000-11248-0 (NYL, HYC, KC) and No.\ R01-2005-000-10352-0
(HWL) and through SRC program Grant CSCMR (KC) and No.\
R11-2000-071 (HWL) and by Korea Research Foundation (KRF) through
Grant No.\ KRF-2005-070-C00044 (NYL, HYC) and No.\
KRF-2005-070-C00055 (HWL), and through BK21 Program (NYL, HYC,
KC, HWL).

\end{acknowledgements}

\section{Appendix: Calculation of singlet and triplet pair functions}
\label{sec:appendix}

We will in this Appendix present the procedure to calculate the
singlet and triplet pair functions using the Usadel formalism
developed in Sec.\ II. As discussed there, the anomalous function
$\bF_S(x,i\w_n)$ in the S region can be written as
 \beqa
\left(\begin{array}{c}
F_S^+(x,i\w_n)\\
F_S^-(x,i\w_n) \end{array}\right) =
 \left(\begin{array}{c}
F_S^+(x,i\w_n)\\
C_S^-(i\w_n)\cosh k_S(x-d_S) \end{array}\right).
\label{eqn:sol_tri1}
 \eeqa
The BC at $x=0$ in terms of ${\bf F}_S$ alone is given by Eq.\
(\ref{eqn:bc_tri_s}).
 \beqa
\xi_S\f{d\bF_S(0)}{dx}=\hat{A}_S\bF_S(0).
 \eeqa
We also have
 \beqa \label{eqn:bc_tri10}
\xi_S\f{d\bF_S(0)}{dx}=\hat{A}\bF_S(0),
 \eeqa
where
 \beqa
{\hat A}=\left(\begin{array}{cc} W(i\w_n) & 0 \\
0 & -A_S \end{array}\right),
 \eeqa
where $W(\w_n)$ and $A_S$ are given by Eq.\ \eqn{eqn:Wwn}. We
therefore write
 \beqa (\hat{A}_s-\hat{A})\bmat
{c}
F_S^+(0,i\w_n)\\
C_S^-(i\w_n)\cosh k_Sd_S \emat = 0.
 \eeqa
Dividing this by $F_S^+$, we obtain
 \beqa  \hat{A}_s \bmat {c} 1 \\ 0  \emat
= \left( \bmat {cc} 1 & 0
\\
0 & -k_S\xi_S \sinh k_Sd_S  \emat  \right.\nonumber \eeqa \beq
\left. -\hat{A_s} \bmat {cc}
0 & 0 \\
0 & \cosh k_Sd_S  \emat\right) \bmat {c} W(i\w_n) \\
C_S^-(i\w_n)/F_S^+(0,i\w_n)
 \emat.
\label{eqn:linear_eq}
 \eeq
Note that $\Delta(x)$ is the eigenvector corresponding to the
smallest eigenvector $M(x,y)$ of Eq.\ (\ref{eqn:Self_TC}), and
$F_S^+(x,i\w_n)$ can be obtained from Eq.\ (\ref{eqn:bi-ss-sol}).
We can therefore find $C_S^-(i\w_n)$ from Eq.\
(\ref{eqn:linear_eq}), and calculate $F_S^-(s,i\w_n)$ using Eq.\
(\ref{sol:fst}). This completes the calculation of the SPC and
TPC in the S region.

We now turn to calculation of the pairing functions in the N
region. The function $F_N^\pm$ can be written as
 \beqa
F_N^\pm(x,i\w_n) = F_N^\pm(0,i\w_n)\cosh k_{N}x \nonumber \\ +
\xi_N\f{d}{dx}F_N^\pm(0,i\w_n) \f{\sinh k_{N}x}{\xi_N k_N}.
 \eeqa
From the BC \eqn{eqn:bc_tri3}, we get
 \beqa
F_N^\pm(x) = F_S^\pm(0)\left[
(1-\f{\g_b^{SN}}{\g_{SN}}\hat{A_S})\cosh k_Nx \right. \nonumber
\\ \left.+
\f{1}{\g_{SN}}\hat {A_S}\f{\sinh k_Nx}{\xi_Nk_N}\right] .
 \eeqa
Therefore we get
 \beqa
F_N^\pm(-d_N) &=& F_S^\pm(0) \left[
(1-\f{\g_b^{SN}}{\g_{SN}}\hat{A_S})\cosh k_Nd_N \right.
\nonumber\\
&-&\left.\f{1}{\g_{SN}}\hat {A_S}\f{\sinh
k_Nd_N}{\xi_Nk_N}\right],
\nonumber \\
 \xi_N\f{d}{dx}F_N^\pm(-d_N) &=& F_S^\pm(0)
\left[\f{1}{\g_{SN}}\hat{A_S}\cosh k_Nd_N \right. \nonumber\\ &-&
\left. \xi_Nk_N (1-\f{\g_b^{SN}}{\g_{SN}}\hat{A_S})\sinh k_Nd_N
\right] .
 \eeqa
$F_N^\pm(-d_N)$ and $\f{d}{dx}F_N^\pm(-d_N)$ is expressed as
$F_S^\pm(0)$ and the results of equation \eqn{eqn:linear_eq} is
used together, we can get $F_N^\pm(x)$.

Next, we calculate $F_F^\pm(x)$ in F. The function $F_F^\pm(x)$
is written as Eq.\ (\ref{eqn:sol_tri_f}). We then use the BC
\eqn{eqn:bc_tri2} to write
 \beqa
\label{eqn:NF_tri1} &&\xi_N\f{d}{dx}F_N^+(-d_N) \nonumber
\\ &&= \g_{NF}\xi_F\left(k_FC_F^+\sinh k_Fd_F +
k_F^*C_F^-\sinh k_F^*d_F \right), \\
\label{eqn:NF_tri2} &&\xi_N\f{d}{dx}F_N^-(-d_N) \nonumber \\
 &&= \g_{NF}\xi_F\left(- k_FC_F^+\sinh k_Fd_F +
k_F^*C_F^-\sinh k_F^*d_F \right).
 \eeqa
We subtract and add \eqn{eqn:NF_tri1} and \eqn{eqn:NF_tri2} to
obtain
 \beqa
C_F^+ = \f{\xi_N\f{d}{dx}F_N^+(-d_N)- \xi_N\f{d}{dx}F_N^-(-d_N)}
{2\g_{NF}\xi_Fk_F\sinh k_Fd_F},\\
C_F^- = \f{\xi_N\f{d}{dx}F_N^+(-d_N)+ \xi_N\f{d}{dx}F_N^-(-d_N)}
{2\g_{NF}\xi_Fk_F^*\sinh k_F^*d_F} .
 \eeqa
We then obtain $F_F^\pm$ by plugging $C_F^\pm$ into Eq.\
(\ref{eqn:sol_tri_f}).
 \beqa &&
F_F^+(x,i\w_n)= \re \left[\f{\xi_N\f{d}{dx}F_N^+(-d_N)
-\xi_N\f{d}{dx}F_N^-(-d_N)}{2\g_{NF}\xi_Fk_F
\tanh k_Fd_F} \right. \nonumber\\
&& \left( \cosh k_F(x+d_N)+ \sinh k_F(x+d_N)\tanh k_Fd_F
\right) \Big], \\
&& F_F^-(x,i\w_n)= -\im \left[ \f{\xi_N\f{d}{dx}F_N^+(-d_N)
-\xi_N\f{d}{dx}F_N^-(-d_N)}{2\g_{NF}\xi_Fk_F
\tanh k_Fd_F}\right.\nonumber \\
&& \left( \cosh k_F(x+d_N) +\sinh k_F(x+d_N)\tanh k_Fd_F \right)
\Big].
 \eeqa
We can therefore calculate the anomalous functions $F$ in
trilayers. The pair functions $\Psi(x)$ are given by
 \beqa
\Psi^\pm(x)=\pi T_c \sum_{\w_n>0}F^{\pm}(x,i\w_n).
 \eeqa
This completes the calculation of the singlet and triplet pairing
components over the entire region of an S/N/F trilayer.

\end{document}